\documentclass[showpacs,preprintnumbers,amsmath,amssymb,superscriptaddress]{revtex4}

\usepackage{graphicx}
\usepackage{dcolumn}
\usepackage{bm}
\usepackage{epstopdf}
\usepackage{mathrsfs}
\usepackage{amssymb}
\usepackage{amsmath}

\def\del{\partial}

\def\pp{\bar{p}}

\def\q{{\boldsymbol q}}

\def\k{{\boldsymbol k}}

\def\bkappa{{\boldsymbol \kappa}}
\def\bbkappa{\bar{\boldsymbol \kappa}}

\def\sM{\text{med}}

\def\bbl{\bar{\boldsymbol{L}}}
\def\bc{\boldsymbol{C}}

\newcommand{\beq}{\begin{equation}}
\newcommand{\eeq}{\end{equation}}
\newcommand{\bes}{\begin{subequations}}
\newcommand{\ees}{\end{subequations}}
\newcommand{\bal}{\begin{align}}
\newcommand{\eal}{\end{align}}
\newcommand{\be}{\begin{eqnarray}}
\newcommand{\ee}{\end{eqnarray}}

\begin{document}

\title{Interference between initial and final state radiation in a QCD medium}

\author{N\'estor Armesto}
\author{Hao Ma}
\author{Mauricio Mart\'{\i}nez}
\affiliation{
Departamento de F\'isica de Part\'iculas and IGFAE,
Universidade de Santiago de Compostela \\
E-15782 Santiago de Compostela, 
Galicia-Spain}
\author{Yacine Mehtar-Tani}
\affiliation{
Institut de Physique Th\'eorique, CEA Saclay, 
F-91191 Gif-sur-Yvette, France
}%
\author{Carlos A. Salgado}
\affiliation{
Departamento de F\'isica de Part\'iculas and IGFAE,
Universidade de Santiago de Compostela \\
E-15782 Santiago de Compostela, 
Galicia-Spain}
\affiliation{Physics Department, Theory Unit, CERN, CH-1211 Gen\`eve 23, Switzerland}
\date{\today}


\begin{abstract}
 We investigate the color coherence pattern between initial and final state radiation in the presence of a QCD medium. We derive the medium-induced gluon spectrum of an ``asymptotic" parton which suffers a hard scattering and subsequently crosses the medium. The angular distribution of the induced gluon spectrum is modified when one includes interference terms between the incoming and the outgoing parton at finite angle between them. The coherent, incoherent and soft limits of the medium-induced gluon spectrum are studied. In the soft limit, we provide a simple and intuitive probabilistic picture which could be of interest for Monte Carlo implementations. The configuration studied here may have phenomenological consequences in high energy nuclear collisions.
\end{abstract}
\maketitle

\section{Introduction}
\label{sec:intro}

Coherence of soft gluon radiation is one of the most important and remarkable properties of perturbative QCD. As an intrinsic quantum phenomenon, coherence arises from destructive interferences leading to a suppression of the spectrum in some regions of the available phase space. A well-known effect is the suppression of large angle radiation that translates into angular ordering of subsequent gluon emissions in a parton cascade and that produces, in particular, the experimentally confirmed \cite{Braunschweig:1990yd,Abbiendi:2002mj} hump-backed plateau of the single inclusive distribution of hadrons inside a jet. 

Among the most important high energy processes where QCD coherence effects play an important role is in the formation of final hadronic states in DIS. In DIS it is necessary to analyze the branching of the incoming parton that is formed long before the hard interaction with the virtual photon and corresponds to a space-like cascade, together with the emissions from the parton outgoing from the hard interaction. The effect of coherence leads to a picture of soft bremsstrahlung emission with angular ordering. Space-like parton branchings determine the behavior of the fragmentation in DIS and the structure of the final hadron states in Drell-Yan and hight-$p_T$ processes, among other observables \cite{Dokshitzer:1991wu,Ellis:1991qj,Bassetto:1984ik}. 

Not much is known about modifications to this vacuum coherence pattern in the presence of a spatially extended QCD medium. Most of the efforts performed during last fifteen years have been concentrated on the medium modifications for the gluon inclusive spectrum off a fast parton without including coherence among different emitters \cite{Baier:1996kr,Baier:1996sk,Zakharov:1996fv,Zakharov:1997uu,Wiedemann:2000za,gyu00}. Recently, an interesting approach to address the problem of coherence effects in the presence of a QCD medium from first principles has been started \cite{MehtarTani:2010ma,MehtarTani:2011tz,MehtarTani:2011gf,Armesto:2011ir,MehtarTani:2011jw,CasalderreySolana:2011rz,MehtarTani:2012cy}.  These works studied  gluon emission off a  $q\bar{q}$ antenna, of opening angle $\theta_{q\bar q}$,  immersed in a QCD medium of length $L$. Different configurations of the antenna being either a color singlet or a color octet state for both the massless \cite{MehtarTani:2010ma,MehtarTani:2011gf} and  the massive case \cite{Armesto:2011ir} for dilute medium, as well as for an opaque medium both in the soft gluon sector \cite{MehtarTani:2011tz} and for finite gluon energies \cite{MehtarTani:2011jw,CasalderreySolana:2011rz,MehtarTani:2012cy}, were considered. 

The emerging picture presents an intriguing and interesting structure of the interference process in the presence of a QCD medium. On the one hand, the medium induces a partial decoherence of the $q\bar q$ antenna which opens the phase space for QCD radiation at large angles, a property which was called {\it anti-angular ordering}  \cite{MehtarTani:2010ma} in the soft limit. This decoherence is total for a dense medium in which the quark and the antiquark emit  independently, even forgetting about their original color configuration being an octet or a singlet. On the other hand, the behavior of the spectrum is defined by the hardest scale in the problem defined as $Q_{hard}= {\rm max}\left\{r_\perp^{-1},m_D\right\}$, so that $Q_{hard}$ indicates the maximum transverse momentum $k_T$ of the emitted gluons. Here $r_\perp=\theta_{q\bar q}L$ is the typical size of the dipole and $m_D$ is the Debye mass of the medium~\footnote{This estimate is valid to first order in the opacity expansion approximation. For the multiple soft scattering case, $\hat q L$ takes the place of $m_D$, with $\hat q$ being the transport coefficient of the medium that characterizes the average transverse momentum squared transferred from the medium to a parton traversing it, per mean free path.}.

In this letter, we extend the studies of coherence effects inside a QCD medium to a different configuration, namely a space-like ($t$-channel) scattering process.  We consider a very simple setup where an "asymptotic" parton created in the remote past suffers a hard scattering and subsequently traverses  a QCD medium of finite size. For simplicity, we limit ourselves in this work to the dilute regime case. We calculate the medium modifications to the inclusive gluon spectrum by including not only the independent gluon emissions of an incoming and outgoing parton but also the interferences between both emitters at finite angle between them. We study and give physical interpretations to the soft, coherent and incoherent limits of the medium-induced gluon spectrum.  

This work is organized as follows: In Sect. \ref{sec:vacinit} we briefly review of the radiation pattern of the initial state radiation in vacuum, using semi-classical methods that is an appropriate framework to derive the gluon spectrum. In Sect. \ref{sec:medgluon} we calculate the medium-induced gluon spectrum for our selected setup. In Sect. \ref{sec:asymlim} we study the coherent, incoherent and soft limit of the medium-induced gluon spectrum. Finally, we present our conclusions and provide an outlook in Sect. \ref{sec:summary}. 


\section{Vacuum emission pattern of radiation}
\label{sec:vacinit}
In this section we briefly describe the vacuum emission pattern of the initial state radiation, restricting ourselves to the color singlet case for simplicity. Thus, we consider a deep inelastic scattering process with a highly virtual photon as the $t$-channel exchange, that we denote {\it hard scattering} in the following. We introduce the semi-classical methods to calculate the scattering amplitudes \cite{Blaizot:2004wu,Blaizot:2004wv,Gelis:2005pt,MehtarTani:2006xq} as well as the notation followed in this work.  In the classical limit, the inclusive spectrum of a radiated on-shell gluon with $4$-momentum $k=(w,\vec{k})$ is given by
\beq
\label{eq:incluspec}
(2\pi)^3 2\omega \ \frac{dN}{d^3\vec{k}}=\sum_{\lambda=\pm 1}|{\cal M}^a_\lambda(\vec{k})|^2 \,.
\eeq
The scattering amplitude is obtained from the classical gauge field through the reduction formula
\beq
\label{eq:redform}
{\cal M}^a_\lambda (\vec{k})=\lim_{k^2\to 0 } -k^2 A^a_\mu (k) \epsilon^\mu_\lambda \,.
\eeq
The classical gauge field, $A_\mu\equiv A^a_\mu t^a$ ($ t^a$ being  the generator of $SU(3)$ in the fundamental representation), is the solution of the classical Yang-Mills (CYM) equations, $[D_\mu,F^{\mu\nu}]=J^\mu$. The covariant derivative is  defined as $D_\mu=\partial_\mu - ig A_\mu$ and the non-abelian field  tensor is $F_{\mu\nu}=\partial_\mu A_\nu-\partial_\nu A_\mu -ig [A_\mu,A_\nu]$. The emitting partons are described by a classical eikonalized current $J^\mu$ which is covariantly conserved i.e. $[D_\mu,J^\mu]=0$. Our calculation is performed in the light cone gauge $A^+=0$. The classical eikonalized current that describes the incoming and outgoing partons after the hard scattering at a certain time $t=t_0$ reads as $J^\mu_{(0)}= J_{in,(0)}^{\mu,a}+J_{out,(0)}^{\mu,a}$, where the currents for the incoming and outgoing quarks are
\beq
\label{eq:inoutcurs}
J_{in,(0)}^{\mu,a}= g\frac{p^\mu}{p^+}\,\delta^{(3)}\biggl(\vec{x}- \frac{\vec{p}}{E} t\biggr) \Theta (t_0-t) Q^{in}_{a}\,,\hspace{0.5cm}
J_{out,(0)}^{\mu,a}= g\frac{\pp^\mu}{\pp^+}\,\delta^{(3)}\biggl(\vec{\bar x}- \frac{\vec{\bar p}}{\bar E} t\biggr) \Theta (t-t_0)
 Q^{out}_{a}\,,
\eeq
with $Q^{in (out)}_{a}$ the color charge of the in(out)coming parton and an overlining $\bar{\ \ }$ is used hereafter to denote the momentum and related quantities of the outgoing quark. By current conservation, one has that $Q^{in}_{a}=Q^{out}_{a}$, and $(Q^{in}_{a})^2=\bigl(N_c^2 -1\bigr)/(2N_c)$.  Due to the gauge choice, it is convenient to perform the calculation in the light cone variables i.e. $k\equiv \bigl[k^+=(\omega+k^3)/\sqrt{2},k^-=(\omega-k^3)/\sqrt{2},\k \bigr]$, $\k=(k^1,k^2)$. By linearizing the CYM equations, the solution of the classical gauge field at leading order in momentum space reads
\beq
\label{eq:solvac}
-k^2 A_{(0)}^{i,a}=\, 2ig \left( \frac{\kappa^i}{\bkappa^2}Q_{in}^a - \frac{\bar\kappa^i}{\bbkappa^2}Q_{out}^a\right) \,,
\eeq
where we introduce the transverse vector $\kappa^i = k^i - x\,p^i$, $i=1,2$. This vector describes the transverse momentum of the gluon relative to the one of the incoming quark (an analogous definition follows for the outgoing quark). In addition, we define the momentum fractions carried out by the emitted gluon with respect to the parent quark, $x$ and $\bar{x}$, as $x= k^+/p^+$ and $\bar{x}=\bar{k}^+/\pp^+$ respectively. 

Taking the solution of the classical gauge field (\ref{eq:solvac}) into the reduction formula (\ref{eq:redform}) and summing over the physical polarizations, the vacuum spectrum of the inclusive gluon spectrum (\ref{eq:incluspec}) for the singlet case results
\beq
\label{vacspec}
\omega\frac{dN^\text{vac}}{d^3\vec{k}}=\frac{\alpha_s  C_F}{\pi^2}\,
\Biggl( \frac{1}{\bkappa^2}+\frac{1}{\bbkappa^2}-2\frac{\bkappa\cdot\bbkappa}{\bkappa^2\bbkappa^2}
\Biggr)
\equiv \frac{\alpha_s\,C_F}{(2\pi)^2\,\omega^2}\,\big(\mathcal{P}^{vac}_{in} + \mathcal{P}^{vac}_{out}\big)  \,,
\eeq
where (an analogous definition follows for $\mathcal{P}^{vac}_{out}$)
\beq
\mathcal{P}^{vac}_{in}= 4\omega^2\Biggl(\frac{1}{\bkappa^2}-\frac{\bkappa\cdot\bbkappa}{\bkappa^2\bbkappa^2}\Biggr)\equiv \mathcal{R}_{in}-\mathcal{J}\,.
\eeq
$\mathcal{R}_{in}$ corresponds to the independent hard emission of the incoming quark and $\mathcal{J}$ is the interference between the two emitters. The inclusive gluon spectrum (\ref{vacspec}) presents soft and collinear divergences. In addition, the spectrum is suppressed at large angles due to destructive interference effects between the emitters. This can be seen clearly by considering inclusive quantities. For instance, by taking the azimuthal average along the longitudinal axis of the incoming quark one obtains that its corresponding gluon emission probability reads   \cite{Dokshitzer:1991wu}
\beq
\label{eq:incvacspec}
\langle dN^{vac}_{in}\rangle_\phi = \frac{\alpha_s \,C_F}{\pi}\,\frac{d\omega}{\omega}\, \frac{d\theta}{\theta}\Theta\bigl(\theta_{qq} - \theta\bigr)\,,
\eeq
where $\theta$ is the angle of emission and $\theta_{qq}$ is the opening angle between the incoming and the outgoing quark which is related with the virtuality $Q^2$ of the off-shell particle (a photon in DIS) in the hard scattering.  Eq. (\ref{eq:incvacspec}) indicates that the  radiated gluon emissions will be confined in the cone with opening angle $\theta_{qq}$ along the incoming or outgoing quark. This property is known as {\it angular ordering}. The procedure described in this section can be extended to higher orders and constitutes the basic building block for the construction of a coherent parton branching formalism \cite{Ellis:1991qj}. 


\section{Radiation in a QCD medium}
\label{sec:medgluon}
\begin{figure}[h]
\begin{center}
\includegraphics[width=0.9\textwidth]{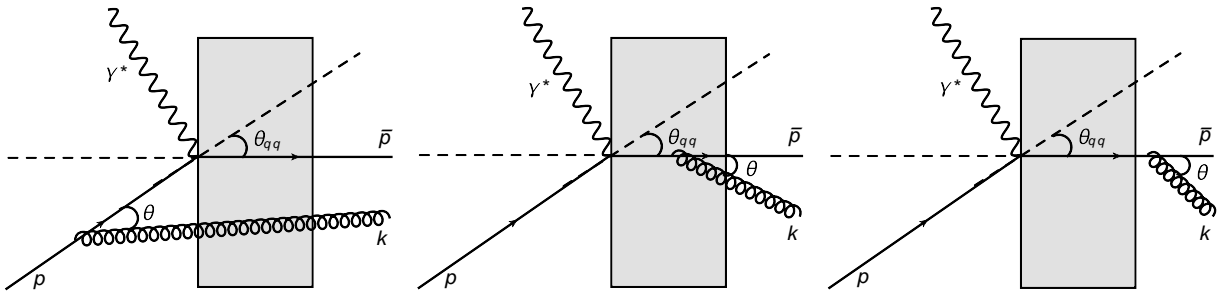}
\end{center}
\caption{Sketch of the $t$-channel scattering process in the presence of the QCD medium. Gluon radiation can take place either before  the hard scattering and the formation of the QCD medium (left) or after  it (middle and right). Note that $\theta_{p\bar p}\equiv \theta_{qq}$ introduced before.}
\label{fig:medsketch}
\end{figure}
In the previous section we gave a short overview of the main aspects of the radiation pattern in the absence of a medium. To include medium modifications to the radiation spectrum, we consider an asymptotic highly energetic quark produced in the remote past that suffers a hard collision at $x^+_0$ and afterwards crosses a static QCD medium of finite size $L^+= \sqrt{2} L$,  $L$ being the size of the medium. For simplicity, the QCD medium is placed exactly after the hard scattering at $x^+=x^+_0$. Gluon radiation is emitted either before or after the hard scattering (see Fig. \ref{fig:medsketch} for an illustration of the physical configuration under consideration).  

In the semi-classical approach, the quark fields act as a perturbation around the strong medium field $A_{med}$ and the total field is written as
\beq
\label{eq:totA}
A^\mu \equiv A^\mu_{med}  + A^\mu_{(0)} + A^\mu_{(1)}\,,
\eeq
where $A_{(0)}$ is the gauge field of the quarks field in vacuum, Eq. (\ref{eq:solvac}), and $A_{(1)}$ is the response of the field at first order in $A_{med}$. 
The medium gauge field is described by $A^-_{med}(x^+,\bold{x})$ and it is a solution of a two-dimensional Poisson equation $-\boldsymbol{\partial} A^-_{med}(x^+,\bold{x})=\rho (x^+,\bold{x})$, where $\rho (x^+,\bold{x})$ is the static distribution of medium color charges that is treated as a Gaussian white noise. Notice that in this approximation $A^i_{med}=A^+_{med}=0$ \cite{MehtarTani:2006xq}. In Fourier space the medium gauge field reads
\beq
\label{eq:Amed}
A^-_{med} (q)=\,2\pi\,\delta(q^+)\,\int_{x^+_0}^\infty dx^+\,\mathcal{A}^-_{med}(x^+,\bold{q})\,e^{i\,q^-x^+}\,.
\eeq
At first order in the medium field, the continuity relation for the induced eikonalized current becomes $\partial_\mu J^\mu_{(1)}=\,ig\,\bigl[A^-_{med},J^+_{(0)}\bigr]$. Its solution can be  written 
\beq
\label{eq:totindcurr}
J^\mu_{(1)}=ig\frac{p^\mu}{p\cdot \del }~ \bigl[A^-_{\sM}, J^+_{in,(0)}\bigr] + ig \frac{ \bar{p}^\mu}{ \bar{p}\cdot \del }~ \bigl[A^-_{\sM}, J^+_{out,(0)}\bigr] 
\equiv J^\mu_{in,(1)} +J^\mu_{out,(1)}\,,
\eeq
where $J^+_{in,(0)}$ and $J^+_{out,(0)}$ are given by the expressions given in Eq. (\ref{eq:inoutcurs}). The current $J^\mu_{in,(1)}$ for the incoming parton is 
\beq
\label{eq:incurrsol}
J^{\mu,a}_{in,(1)}(k)=(ig)^2\frac{p^\mu}{-i(p\cdot k) }\int\!\! \frac{d^4q}{(2\pi)^4}~\frac{p^+}{p\cdot (k-q)-i\epsilon}~i[T\cdot A^-_{\sM}(q)]^{ab} Q_{in}^b \,,
\eeq
where $[T\cdot A^-_{\sM}(q)]^{ab} Q_{in}^b \equiv -i f^{abc}A^c_{\sM}(q)Q^b_{in}$ and $f^{abc}$ is the $SU(3)$ structure constant. An analogous expression is found for the current $J^\mu_{out,(1)}$ for the outgoing quark. 

After linearizing the CYM equations, the equation of motion for the transverse components of the induced gauge field $A^i_{(1)}$ is
\beq
\label{eq:A1eq}
\square A_{(1)}^i-2ig  \bigl[A^-_{\sM}, \, \del^+ A_{(0)}^i\bigr] = -\frac{\del^i}{\del^+}J_{(1)}^++J_{(1)}^iÊ\,.
\eeq

Now we  focus in the contribution to the gluon emission scattering amplitude. For completeness, we calculate explicitly the contribution of the incoming quark. We omit the calculation of the amplitude associated to the outgoing quark whose details can be found in Refs.\cite{MehtarTani:2010ma,MehtarTani:2011gf}. As we shall show below,  the radiative gluon emission associated to the outgoing quark turns out to be different to the one associated to the incoming quark.

The solution of Eq. (\ref{eq:A1eq}) in Fourier space for the induced gauge field due the incoming quark, $A^-_{in,(1)}$,  can be written as
\beq
\label{eq:Ain1sola}
-k^2 A^-_{in,(1)} = 2g \int \frac{d^4q}{(2\pi)^4}\, (k-q)^+ \bigl[A_{\sM}^-(q), A^i_{in,(0)}(k-q)\bigr] -\frac{k^i}{k^+}J^+_{in,(1)}+J^i_{in,(1)} \,,
\eeq
where $A^i_{in,(0)}(k)$ is identified with the incoming quark-induced vacuum field from Eq. (\ref{eq:solvac}),
\beq
-k^2A^{i,a}_{in,(0)}(k)\;=\;- 2ig\frac{\kappa^i}{\bkappa}\,Q_{in}^a \;.
\eeq
Integrating out $q^-$ in Eq. (\ref{eq:Ain1sola}) and imposing the condition that the medium starts at $x^+=x^+_0$, one obtains
\beq
\label{eq:Ain1solb}
-k^2A_{in,(1)}^{i,a} = 2ig^2  \int \frac{d^2 \q}{(2\pi)^2}\,\int^{+ \infty}_{x^+_0} d x^+\, [T\cdot A^-_{\sM}(x^+,\q)]^{ab} Q_{in}^b\,e^{i \left(k^--\frac{(\k-\q)^2}{2 k^+}\right)x^+} \frac{(\kappa-q)^i}{\bigl(\bkappa-\q\bigr)^2}  \,.
\eeq
On the other hand, the solution of the induced gauge field due to the outgoing quark reads \cite{MehtarTani:2010ma,MehtarTani:2011gf}
\be
\label{eq:Aout1sol}
-k^2A_{out,(1)}^{i,a}& = &- 2ig^2  \int \frac{d^2 \q}{(2\pi)^2}\,\int^{+ \infty}_{x^+_0} d x^+\, [T\cdot A^-_{\sM}(x^+,\q)]^{ab} Q_{out}^b\, e^{i \left(k^--\frac{(\k-\q)^2}{2 k^+}\right)x^+} \nonumber \\
&\times& \left[\frac{(\bar{\kappa}-q)^i}{(\bbkappa-\q)^2}\left\{1-\exp\biggl(i \frac{(\bbkappa-\q)^2}{2 k^+}x^+\biggr)\right\}+\frac{\bar{\kappa}^i}{\bbkappa^2} \exp\biggl(i \frac{(\bbkappa-\q)^2}{2 k^+}x^+\biggr) \right] \,.
\ee
Thus,  using the reduction formula (\ref{eq:redform}) together with Eqs. (\ref{eq:Ain1solb}) and (\ref{eq:Aout1sol}) we get the total scattering amplitude for gluon radiation off the incoming and outgoing quark:
\be
\label{eq:totmedampl}
{\cal M}^a_{\lambda} &=& {\cal M}^a_{\lambda,in} +{\cal M}^a_{\lambda,out}\nonumber\\
&=&2ig^2  \int \frac{d^2 \q}{(2\pi)^2}\,\int^{L^+}_{x_0^+} d x^+\, [T\cdot A^-_{\sM}(x^+,\q)]^{ab}\nonumber\\
&\times&\biggl\{
Q_{in}^b\,\frac{\bkappa-\q}{\bigl(\bkappa-\q\bigr)^2} 
-Q_{out}^b\, \left[\frac{\bar{\bkappa}-\q}{(\bbkappa-\q)^2}-\bbl \exp\biggl(i \frac{(\bbkappa-\q)^2}{2 k^+}x^+\biggr) \right] \biggr\}\,,
\ee
modulo a phase that cancels in the cross section. Here we use the definition of the transverse component of the Lipatov vertex in the light cone gauge:
\beq
\label{eq:lipvertex}
\bbl=\frac{\bar{\bkappa}-\q}{(\bbkappa-\q)^2}-\frac{\bar{\bkappa}}{\bbkappa^2} \,.
\eeq

We point out that the contribution of the incoming and the outcoming parton to the total gluon emission scattering amplitude is not the same. The difference between both contributions is due to the fact that gluons emitted by the outcoming parton are induced by the presence of the QCD medium, while gluons radiated by the incoming parton are created before the hard scattering and its encountering with the medium - thus the observed net effect after crossing the medium is the reshuffling of their momenta. 

To evaluate the cross section we must average over the medium field and include the virtual corrections - the so-called contact terms \cite{Baier:1996kr,Baier:1996sk, Wiedemann:2000za}. To perform the medium average, we assume the usual approximation for the case of a single scattering center
\beq
\label{eq:MediumAverage}
\langle A^-_{\sM,a}(x^+,\q) A^{- *}_{\sM,b}(x'^+,\q')\rangle = \delta^{ab} m_D^2~ n_0~\delta(x^+-x'^+)(2\pi)^2 \delta^{(2)}(\q-\q') \, {\cal V}^ 2(\q) \,,
\eeq
where $ {\cal V}(\q)=1/(\q^2+m_D^2)$ is a Yukawa-type potential with a Debye mass $m_D$ and $n_0$ is the one-dimensional density that describes the distribution of the scattering centers along the QCD medium. The contact terms required by unitarity are the interferences between the gluon emission amplitude in vacuum and the one accompanied by two scatterings with the medium with no net momentum transfer. These can be added to the radiative cross section through a redefinition of the potential ${\cal V}^2(\q)\to{\cal V}^2(\q)-\delta^2(\q)\int d^2\q' {\cal V}^2(\q')$ \cite{Wiedemann:2000za}.

\begin{figure*}[t]
\label{fig:gluontag}
\begin{center}
\includegraphics[width=12cm]{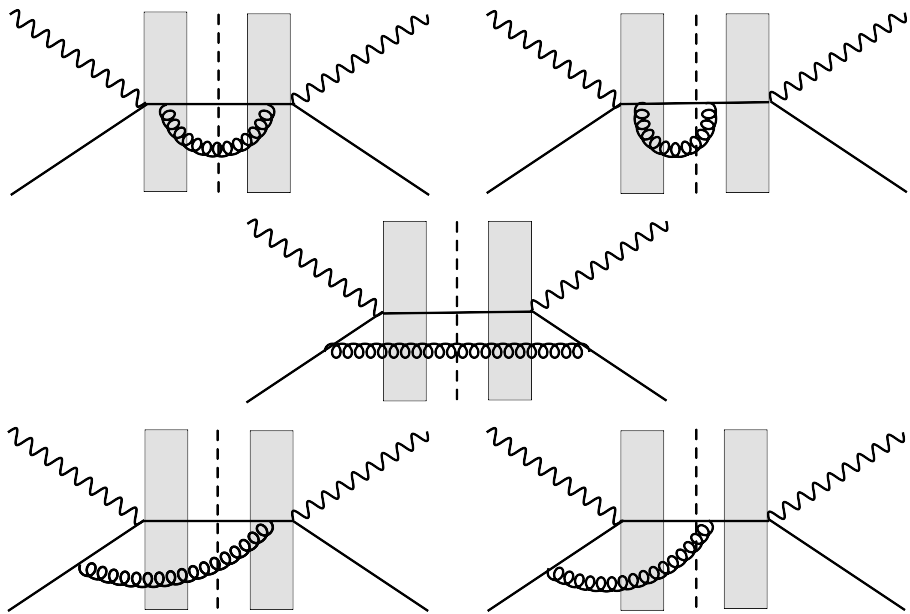}
\end{center}
\caption{Different classes of Feynman diagrams which contribute to the radiative cross section in light-cone gauge: independent gluon emissions from the outgoing quark (top), from the incoming quark (middle) and  interferences (bottom). See text for details.}
\label{fig:gluoncases}
\end{figure*}

When calculating the radiative cross section in the light-cone gauge, the contributions to the gluon spectrum can be separated according to the position of the gluon emission vertex in the amplitude and its complex conjugate with respect to the photon vertex - the hard scattering. In our setup, we have three possible cases: (i) the independent gluon radiation off the outgoing quark, which corresponds to the square of the amplitudes with both emission vertices after the hard scattering; (ii) the independent gluon radiation off the incoming quark, which corresponds to the square of the amplitude where the emission vertex is before the hard scattering, so that the gluon entering the medium is fully formed; (iii) the inferences between the cases where the gluon vertex is before and after the hard scattering in amplitude and complex conjugate of the amplitude. Fig. 2 illustrates this classification in terms of Feynman diagrams \footnote{Note that, as shown in detail in previous works \cite{Armesto:2011ir}, the same result as presented here can be obtained using the Feynman diagram language from the start instead of the semi-classical method. In this case, there are 2 vacuum amplitudes ${\cal M}_0$, 5 amplitudes with one scattering ${\cal M}_1$ and 7 amplitudes with two scatterings that contribute to the contact terms ${\cal M}_2$. The medium-induced gluon spectrum is proportional to the sum of medium averages $\langle|{\cal M}_1|^2\rangle + 2 {\bf Re}\langle{\cal M}_0^\dagger {\cal M}_2\rangle$ which consists of 39 diagrams.}.

After squaring the amplitude and summing over the polarization vectors and considering the interferences, the spectrum of the medium-induced gluon radiation reads
\be
\label{eq:totmedspec}
\omega\frac{dN^\text{med}}{d^3\vec{k}} &=&\frac{4\,\alpha_s C_F \,\hat q}{\pi} \int \frac{d^2 \q}{(2\pi)^2}{\cal V}^2(\q)~\int_0^{L^+} dx^+\,\Biggl[\frac{1}{(\bkappa-\q)^2}-\frac{1}{\bkappa^2}\nonumber\\
&+&\,2\,\frac{\bbkappa\cdot\q}{\bbkappa^2(\bbkappa-\q)^2}\biggl(1-\cos\biggr[\frac{(\bbkappa-\q)^2}{2 k^+}x^+\biggr]\biggr)\,\nonumber\\
&+&\,2\,\Biggl\{\frac{\bkappa\cdot\bbkappa}{\bkappa^2\bbkappa^2}-\frac{\bbkappa\cdot(\bkappa-\q)}{\bbkappa^2(\bkappa-\q)^2}\Biggr\}
\,+\, 2 \Biggl\{\frac{\bbkappa\cdot(\bkappa-\q)}{\bbkappa^2(\bkappa-\q)^2} -\frac{(\bbkappa-\q)\cdot(\bkappa-\q)}{(\bbkappa-\q)^2(\bkappa-\q)^2}\,
\,\Biggr\}\biggl(1-\cos\biggr[\frac{(\bbkappa-\q)^2}{2 k^+}x^+\biggr]\biggr)
\Biggr]\,,
\ee
where we define $\hat{q}=\alpha_s C_A n_0 m_D^2$ and  set $x_0^+=0$. The first line in Eq. (\ref{eq:totmedspec}) is the contribution to the gluon radiation spectrum  off  the incoming quark which corresponds to the bremsstrahlung of the accelerated charge which subsequently undergoes  rescattering. The second line in Eq. (\ref{eq:totmedspec}) corresponds to emissions off the outgoing quark; it is identified with the so-called Gyulassy-Levai-Vitev (GLV) spectrum \cite{Wiedemann:2000za,gyu00} or equivalently, the first order in opacity of the Baier-Dokshitzer-Mueller-Peign\'e-Schiff-Zakharov-Wiedemann spectrum \cite{Baier:1996kr,Baier:1996sk,Zakharov:1996fv,Zakharov:1997uu,Wiedemann:2000za}. The novel contributions associated to interferences between both emitters are contained in the third line. Note that the interference terms show diverse phase structures, and that the interference contrbutions shows soft and collinear divergencies. 

We observe that in the limit when the opening angle between both emitters vanishes ($\theta_{qq}\to 0$), the medium-induced gluon spectrum (\ref{eq:totmedspec}) reduces to the well known  Gunion-Bertsch (GB) result \cite{Gunion:1981qs}. This is expected since the GB spectrum is the genuine induced emission due to a hard scattering of an asymptotic charge with the medium. Other important asymptotic limits of the gluon spectrum (\ref{eq:totmedspec}) are the incoherent, coherent and soft limit which we shall consider in the next section. 


\section{Asymptotic limits}
\label{sec:asymlim}

In the previous section, we calculate the medium-induced gluon spectrum (\ref{eq:totmedspec}) by taking into account the independent  emissions off either the incoming or outgoing parton as well as the interferences between them. The pattern of interferences between hard \footnote{In our notation, {\it hard emission} is the one in which the emitted gluon undergoes no scattering with the QCD medium.} and medium-induced emissions is complex so, to better elucidate its physics, we consider in this section three interesting asymptotic regimes, namely the incoherent, coherent and soft limits. 


\subsection{Incoherent limit}
\label{subsec:incohlimit}

The GLV spectrum shows an interplay between hard  and medium-induced gluon emissions. The phase achieved by color rotation relates the effective formation time of the emitted gluon $\tau_f\sim \omega/(\bbkappa-\q)^2$ to the position of the interaction with the medium which is the main cause of interference between both mechanisms of emission. The spectrum reaches its maximum  in the incoherent limit $\tau_f \ll L^+$ where these mechanisms can be clearly separated since the phases cancels. In addition, this limit permits a clear probabilistic interpretation \cite{Wiedemann:2000za}. 

When one takes the incoherent limit, $\tau_f \ll L^+$ , the  contribution with the cosines in Eq. (\ref{eq:totmedspec}) can be neglected and the spectrum can be written
\beq
\label{eq:incohlimit}
\omega\frac{dN^\text{med}}{d^3\vec{k}}\biggl.\biggr|_{\tau_f\ll L^+} =\frac{4\,\alpha_s C_F \,\hat q}{\pi} \int \frac{d^2 \q}{(2\pi)^2}{\cal V}^2(\q)~\int_0^{L^+} dx^+ \biggl\{\bbl^2+\bc^2(\bkappa-\q)-\bc^2(\bkappa) \biggr\}\,,
\eeq
where we use the definition of the transverse emission current (an identical definition follows for $\bc (\bkappa-\q)$ by changing $\bkappa\to\bkappa-\q$)
\beq
\bc (\bkappa)= \frac{\bkappa}{\bkappa^2}-\frac{\bbkappa}{\bbkappa^2}\,.
\eeq
Notice that the gluon spectrum in vacuum, Eq. (\ref{vacspec}), can be rewritten in terms of $\bc (\bkappa)^2$ so this term takes into account the independent hard gluon emissions as well as their interferences. A similar interpretation applies for $\bc (\bkappa-\q)^2$ but by reshuffling the momenta of the emitted gluon due to the interaction with the medium. 

The incoherent limit of the gluon spectrum given by Eq.  (\ref{eq:incohlimit}) allows a similar probabilistic interpretation as in the case of the GLV spectrum but now including interferences. In this case, we observe two mechanisms of gluon radiation: The first term of Eq. (\ref{eq:incohlimit}) corresponds to the genuine medium-induced radiation of an asymptotic parton that suffers a scattering with the medium (GB). The two last terms corresponds to the the radiation pattern (\ref{vacspec}) (bremsstrahlung associated to the hard scattering)  followed by the radiated gluon suffering a classical  sequential process of rescattering with the medium. In addition, these two last terms include interferences among the incoming and the outgoing quarks alike. 


\subsection{Coherent limit}
\label{subsec:cohlimit}

For large formation times $\tau_f\gg L^+$, the medium-induced gluon spectrum off a parton created at finite time is completely suppressed due to the LPM effect.
When this limit is taken for the gluon spectrum (\ref{eq:totmedspec}), the cosine does not oscillate so $\cos (L^+/\tau_f)\to 1$ and the gluon spectrum is further simplified to

\beq
\omega\frac{dN^\text{med}}{d^3\vec{k}}\biggl.\biggr|_{\tau_f \gg L^+}= \frac{4\,\alpha_s C_F \,\hat q}{\pi} \int \frac{d^2 \q}{(2\pi)^2}{\cal V}^2(\q)~\int_0^{L^+} dx^+\biggl\{
\frac{1}{(\bkappa-\q)^2}-\frac{1}{\bkappa^2} 
+ 2\frac{\bbkappa\cdot\bkappa}{\bbkappa^2\bkappa^2}
-2 \frac{\bbkappa\cdot(\bkappa-\q)}{\bbkappa^2(\bkappa-\q)^2} 
\biggr\}\,.
\label{eq:coherent}
\eeq
The first two terms in (\ref{eq:coherent}) correspond simply to the reshuffling of the gluon emission off the incoming quark. The next two terms are interferences between the initial and final state. The fact that there is no contribution exclusively associated to the outgoing parton (GLV) is caused mainly by the destructive interferences between hard emissions and medium-induced emissions taking place completely inside the medium. So, in this case, the LPM effect becomes subleading and some of the interferences among the initial and final quarks remain \footnote{Notice that when Eq. (\ref{eq:coherent}) is integrated in $k_\perp$, neglecting finite kinematics of the gluon, one recovers the result shown in Eq. (3.4) of \cite{Arleo:2010rb} where a similar setup as the one presented here has been advocated to be of relevance for energy loss of quarkonia in nuclear matter.}. In the coherent limit the position of the scattering center become 'close' to the hard production point where the medium starts and, thus, it looks as if the hard gluon radiated from the outgoing parton does not rescatter and it is produced completely outside the medium.  This hard gluon produced outside the medium is precisely the one that interferes with any of the hard and medium-induced gluon emissions associated to the incoming quark. 


\subsection{Soft limit}
\label{subsec:softlimit}

 The medium-induced gluon spectrum (\ref{eq:totmedspec}) can be further simplified in the soft limit $\omega\to 0$. In this limit,  the dominating contributing terms of the radiative scattering amplitude $|{\cal M}|^2$ are
\beq
\label{eq:softlim-ampl}
\lim_{\omega\to0} |{\cal M}|^2= \frac{4\,\alpha_s\, C_F \,\hat q}{\pi}  \biggl(2\frac{\bkappa\cdot\bbkappa}{\bkappa^2\,\bbkappa^2}-\frac{1}{\bkappa^2}\biggr)\,.
\eeq
From this result, we conclude that in the soft limit medium-induced gluon emissions contain the same structures as the vacuum  ones, since the first term is identified with ${\cal J}/(4\omega^2)$ while the second one corresponds to ${\cal R}_{in}/(4\omega^2)$ (see their definitions in Eq. (\ref{vacspec})). Thus, in the soft limit, the medium-induced gluon spectrum can be written
\beq
\label{eq:softlimspec}
\omega\frac{dN^\text{med}}{d^3\vec{k}}\biggl.\biggr|_{\omega\to0} 
=\frac{\alpha_s C_F }{(2\pi)^2}\,\Delta_{med}\, \bigl(2{\cal J}-{\cal R}_{in}\bigr)\,,
\eeq
where $\Delta_{med}=\hat{q}L^+/m_D^2\approx L/\lambda$ that is the opacity expansion parameter i.e. the effective number of scattering centers. 

In the soft limit, the full gluon spectrum in the presence of a medium is $dN^{tot}|_{\omegaº\to 0}=(dN^{vac}+dN^{med})|_{\omega\to 0}$. Following a similar procedure as indicated in Sect. II, we separate the independent and the interference contributions to the incoming and the outgoing quark in such a way that the full gluon spectrum reads 
\beq
\label{eq:fullspec}
\omega\frac{dN^\text{tot}}{d^3\vec{k}}\biggl.\biggr|_{\omega\to0} 
= \frac{\alpha_s C_F }{(2\pi)^2}\,\bigl({\cal P}_{in}^{tot}+{\cal P}_{out}^{tot}\bigr)\bigr|_{\omega\to0}\,,
\eeq
where 
\bes
\be
\label{eq:totcontra}
{\cal P}_{in}^{tot}&=& \bigl(1-\Delta_{med} \bigr) \bigl( {\cal R}_{in} - {\cal J} \bigr) \,,\\
{\cal P}_{out}^{tot}&=&{\cal R}_{out} \,-\,\bigl(1-\Delta_{med}\bigr)\, {\cal J}\,. \label{eq:totcontrb}
\ee
\ees
In the absence of a medium, $\Delta_{med}\to 0$, one recovers the radiation pattern observed in vacuum, Eq.(\ref{vacspec}). In the opposite case of an opaque medium $\Delta_{med}\to 1$, there is a significant reduction of soft gluon radiation from the incoming parton which is in qualitative agreement with expectations from the saturation of partonic  densities \cite{Iancu:2003xm}. Notice that the bound $\Delta_{med}\le 1$, although not evident in the first order in the opacity expansion, is given by unitarity. When multiple scatterings are considered we expect the soft gluon emission from the incoming quark to be exponentially suppressed \cite{forthcom}.

In the soft limit, the full gluon spectrum (\ref{eq:fullspec}) presents both soft and collinear divergences. The splitting of the gluon radiation into two components, Eqs. (\ref{eq:totcontra}) and (\ref{eq:totcontrb}), allows  not only a separation of the collinear divergences but also a simple and intuitive probabilistic interpretation. The first term ${\cal P}_{in}^{tot}$ is the gluon emission off the incoming quark reduced by the probability $\Delta_{med}$ that an interaction of the emitted gluon occurs inside the medium.  After performing an azimuthal angle average, ${\cal P}_{in}^{tot}$ has the same angular ordering constraint as in the vacuum case i.e.  the emitted gluons will be confined inside a cone with opening angle $\theta_{qq}$ around the incoming quark but soft gluon radiation decreases by a quantity proportional to $\Delta_{med}$. The second term ${\cal P}_{out}^{tot}$ accounts for the partial decoherence of the emitted gluon due to the scatterings with the colored medium.  Such decoherence is measured by the probability of an interaction with the medium $\Delta_{med}$.  ${\cal P}_{out}^{tot}$ shows resemblance to the radiation pattern already observed for the antenna in the dilute regime case \cite{MehtarTani:2010ma,MehtarTani:2011gf}: the interaction with the medium opens the phase space for large angle emissions and there is a strict geometrical separation between vacuum and medium-induced radiation, a property called antiangular ordering there \cite{MehtarTani:2010ma,MehtarTani:2011gf}. 

The soft limit remains valid for a range of finite gluon energies as far as $\omega\theta_{qq},\k\ll m_D$. Hence, the analysis for finite values of  gluon energy involves two regimes related with the relevant scales of the problem, the opening angle $\theta_{qq}$ between the incoming and the outgoing parton and the typical transfer momentum $|\q|\sim m_D$ to the medium. So one has either $\theta_{qq}\lesssim m_D/\omega$ or $\theta_{qq}\gtrsim m_D/\omega$. Numerical and analytical studies for finite gluon energies in the dilute regime case will be presented in a separate publication \cite{forthcom2}.


\section{Summary and outlook}
\label{sec:summary}

In this work we have studied the interference pattern between  initial and final radiation in a QCD medium. We derive an analytic expression for the medium-induced gluon spectrum of an asymptotic parton created in the remote past which suffers a hard collision and subsequently crosses a QCD medium. The medium-induced gluon spectrum has three contributions: the independent gluon emissions associated to the incoming and outgoing parton, and the interference terms between both emitters. The angular distribution of gluon emission results affected by the presence of these interferences between the emitters when one compares with the radiation pattern in the vacuum.

We examine the incoherent, coherent and soft limits of the medium-induced gluon spectrum. For the incoherent limit $\tau_f\ll L^+$, the gluon spectrum allows a similar probabilistic interpretation to the spectrum of a parton formed at finite time (GLV), but now it becomes generalized by including interferences among the two emitters. In the coherent limit $\tau_f\gg L^+$, the GLV contribution is cancelled due to the LPM effect, but some of the interferences remain as well as the classical broadening contribution associated exclusively to the initial state.  In the soft limit, the radiation pattern of the full gluon spectrum retains certain vacuum-like structures which enable a simple and  intuitive probabilistic interpretation. In this soft limit, the two main features of the full gluon spectrum are: (i)  gluon emissions from the initial state are reduced by a quantity which depends on the medium properties $\Delta_{med}\sim \hat{q}/m_D^2$ but, as in vacuum, gluon radiation is confined inside the cone defined by the opening angle $\theta_{qq}$ along the incoming quark and (ii) the final state emissions lose totally their vacuum coherence characteristics once in the medium i.e. large angle emissions (antiangular ordering) arise  from the medium-induced coherent radiation between both emitters. Similar properties for the final state were already pointed out for the $q\bar{q}$ antenna in the dilute regime case \cite{MehtarTani:2010ma,MehtarTani:2011gf}. 

The setup studied in this work shows in a transparent manner how the interferences affect the angular distribution of the gluon radiation under the presence of a QCD medium. 
Note that, as previous studies in vacuum and medium, the results presented here for the $t$-channel exchange of a color singlet object hold, in the soft limit, for arbitrary color representations. In this way, they are applicable to medium-induced soft gluon production for any $t$-channel hard scattering.  
The possible phenomenological consequences of our findings can be considered at two levels: First, at the level of the inclusive spectra of produced gluons, the opening of phase space for large angle soft emissions - evident in the soft limit -  should lead to an increase of soft hadron multiplicities. Second and at a more fundamental level, the presence of non-vanishing interferences between initial and final state radiation questions the validity of usual collinear factorization in high-energy collisions involving a QCD medium. Even in the soft limit where factorization seems to be restored, the results (\ref{eq:fullspec}), (\ref{eq:totcontra}), (\ref{eq:totcontrb}) are strongly suggestive of a modification of the quantum evolution of partonic densities inside nuclei. 
Therefore, this study constitutes a starting point for the proper inclusion of initial state radiation effects and the study of coherence effects on observables sensitive to  initial state radiation in proton-nucleus, nucleus-nucleus or lepton-nucleus collisions, as well as for the improvement of Monte Carlo  models for particle production in collisions involving spatially extended QCD media.
All these aspects deserve further investigation that we plan to address in the future.


\begin{acknowledgments}
The authors would like to thank to A. H. Mueller, G. Beuf, J. G.~Milhano and K.~Tywoniuk for useful discussions. This work is supported by European Research Council grant HotLHC ERC-2001-StG-279579; by Ministerio de Ciencia e Innovaci\'on of Spain under projects FPA2008-01177, FPA2009-06867-E and FPA2011-22776; by Xunta de Galicia (Conseller\'{\i}a de Educaci\'on and Conseller\'\i a de Innovaci\'on e Industria -- Programa Incite); by the Spanish Consolider-Ingenio 2010 Programme CPAN and by FEDER. CAS is a Ram\'on y Cajal researcher. The work of Yacine Mehtar-Tani is supported by the European Research Council under the Advanced Investigator Grant ERC-AD-267258.
\end{acknowledgments}
 

\end{document}